\documentclass[showpacs,amsmath,amssymb]{revtex4}

\usepackage{graphicx}
\usepackage{dcolumn}
\usepackage{bm}
\usepackage{epsfig}

\begin{document}

\def\0#1#2{\frac{#1}{#2}}
\def\bct{\begin{center}} \def\ect{\end{center}}
\def\beq{\begin{equation}} \def\eeq{\end{equation}}
\def\bea{\begin{eqnarray}} \def\eea{\end{eqnarray}}
\def\nnu{\nonumber}
\def\n{\noindent} \def\pl{\partial}
\def\g{\gamma}  \def\O{\Omega} \def\e{\varepsilon} \def\o{\omega}
\def\s{\sigma}  \def\b{\beta} \def\p{\psi} \def\r{\rho}
\def\G{\Gamma} \def\S{\Sigma} \def\l{\lambda}

\title{Bose-Einstein condensation in linear sigma model at Hartree and large $N$ approximation}
\author{Song~Shu}
\affiliation{Faculty of Physics and Electronic Technology, Hubei
University, Wuhan 430062, China}
\author{Jia-Rong~Li}
\affiliation{Institute of Particle Physics, Hua-Zhong Normal
University, Wuhan 430079, China}
\begin{abstract}
The BEC of charged pions is investigated in the framework of O(4)
linear sigma model. By using Cornwall-Jackiw-Tomboulis formalism,
we have derived the gap equations for the effective masses of the
mesons at finite temperature and finite isospin density. The BEC
is discussed in chiral limit and non-chiral limit at Hartree
approximation and also at large $N$ approximation.
\end{abstract}  \maketitle

\section{Introduction}
In recent years a isospin chemical potential has been introduced
into the study for QCD phase structure~\cite{ref1,ref2,ref3}. It
allows one to have a new dimension to study the rich phases of QCD
theory. When compared to the baryon chemical, this isospin
chemical potential in principle has no fermion sign problem on a
lattice simulation~\cite{ref1}. From recent lattice calculation
and the investigation of chiral perturbation theory at finite
isospin chemical potential, it has been indicated that there will
be a pion condensation or pion superfluid phase~\cite{ref4,ref5}.
At finite temperature and finite isospin density it is in essence
a Bose-Einstein condensation (BEC) of charged pion in momentum
space. This kind phenomenon has been also studied by using NJL
model~\cite{ref6,ref7,ref8} and ladder QCD~\cite{ref9}.

In the low energy effective models, the linear sigma model is very
simple and illustrative. When only considering the mesonic part of
the model, there are only four scalar fields, the sigma field and
the usually three pion fields which display a O(4) symmetry. The
model is so called O(4) linear sigma model. It has been usually
used to study the chiral phase transition and it is also well
suited for describing the physics of
meson~\cite{ref10,ref11,ref12}. This model has been well studied
at finite temperature within the Cornwall-Jackiw-Tomboulis (CJT)
formalism~\cite{ref10}. In our previous work, we have introduced
the isospin chemical potential into the linear sigma model and
discussed the BEC of pions and its relation to the chiral phase
transition in the chiral limit within the same
formalism~\cite{ref13}. In this paper, we will extend this work to
discuss the BEC not only in chiral limit but also in non-chiral
limit at the Hartree approximation and also at the large $N$
approximation. Through this work we wish to give a clearer picture
of BEC in linear sigma model at finite temperature and finite
isospin density.

The organization of this paper is as follows. In section 2 we give
a brief introduction of the O(4) linear sigma model, then we
introduce the isospin chemical potential. The CJT formalism will
be briefly described based on $\l\phi^4$ theory. In section 3, the
CJT method is used to derive the gap equations and thermodynamic
functions, then we will discuss the BEC in chiral limit and
non-chiral limit at Hartree approximation. In section 4, the BEC
is discussed at large $N$ approximation. The last section is the
summary.

\section{The linear sigma model at finite isospin chemical potential and the CJT formalism}
We start our discussion from the Lagrangian of the linear sigma
model with only the mesonic part presented, \bea {\cal
L}=\012(\pl\bf\s)^2&+&\012(\pl\vec\pi)^2-\012m^2\s^2-\012m^2\vec\pi^2
\nnu
\\ &-&\0\l{24}(\s^2+{\vec\pi}^2)^2+\e\s , \eea
where $\s$ and ${\vec\pi}$ are the sigma field and the three pion
fields ($\pi_1, \pi_2, \pi_3$) respectively. $\e\s$ is the
explicit chiral symmetry breaking term, where $\e=f_\pi m^2_\pi$
and $f_\pi=93MeV$ is the pion decay constant. At tree level and
zero temperature the parameters of the Lagrangian are fixed in the
way that the masses agree with the observed value of pion mass
$m_\pi=138MeV$ and the most commonly accepted value for sigma mass
$m_\s=600MeV$. Then the coupling constant $\l$ and negative mass
parameter $m^2$ of the model are chosen to be
$\l=3(m^2_\s-m^2_\pi)/f^2_\pi$ and $-m^2=(m^2_\s-3m^2_\pi)/2>0$.
When $\e=0$, the chiral symmetry is spontaneously broken, and pion
is the Goldstone boson which mass is $m_\pi=0$ at zero
temperature.

The isospin chemical potential can be introduced by different
means. In Ref.~\cite{ref14}, the chemical potential is introduced
according to the covariant way fixed by the gauge invariance. In
our previous work~\cite{ref13}, the chemical potential is
introduced through the conserved isospin charge. Both give the
identical results. If we redefine the pion fields as \bea
\pi_-\equiv\01{\sqrt2}(\pi_1+i\pi_2), \ \
\pi_+\equiv\01{\sqrt2}(\pi_1-i\pi_2),\ \ \pi_0\equiv\pi_3. \eea
the Lagrangian with the isospin chemical potential $\mu$ included
can be written as \bea\label{lag} {\cal
L}=\012(\pl_\mu\s)(\pl^\mu\s)+\012(\pl_\mu\pi_0)(\pl^\mu\pi_0)+(D_\mu\pi)^+(D^\mu\pi)^--V(\s,\vec\pi),
\eea where the potential is \bea
V(\s,\vec\pi)=\0{m^2}2(\s^2+\vec\pi^2)+\0\l{24}(\s^2+\vec\pi^2)-\e\s,
\eea and $(D_0\pi)^\pm=(\pl_0\pm i\mu)\pi_\pm$,
$(D_i\pi)^\pm=\pl_i\pi_\pm$ for $i=1,2,3$.

By shifting the sigma field as $\s\to\s+\phi$, where $\phi$ is the
expectation value of the sigma field and it is also the order
parameter of the chiral phase transition, the classical potential
takes the form \bea U(\phi)=\012m^2\phi^2+\0\l{24}\phi^4-\e\phi,
\eea while the interaction Lagrangian which describes the vertices
of the shifted theory is given by \bea {\cal
L}_{int}=-\0\l{12}\s^2\vec\pi^2-\0\l{24}\s^4-\0\l{24}\vec\pi^4-\0\l6\phi\s\vec\pi^2-\0\l6\phi\s^3.
\eea

By a generating functional theory, one can obtain the effective
potential based the above Lagrangian. At finite temperature the
effective potential is identical to the thermodynamic potential
which is very important in discussing the thermodynamic properties
of the system. The usually effective potential $V(\phi)$ depends
on $\phi$, a possible expectation value of the quantum field
$\Phi$. A generalized effective potential for composite operators
has been introduced by Cornwall, Jackiw and
Tomboulis(CJT)~\cite{ref15}. According to this formalism, the
effective potential $V(\phi, G)$ depends not only on $\phi$, but
also on $G(x,y)$, a possible expectation value of the time-ordered
product $T\Phi(x)\Phi(y)$, which is also the full propagator of
the field. Physical solutions demand minimization of the effective
potential with respect to both $\phi$ and $G$, which means
\beq\label{mini} \0{dV(\phi,G)}{d\phi}=0, \ \ \ \ \ \
\0{dV(\phi,G)}{dG}=0. \eeq The derivations to $\phi$ and $G$ are
functional. This formalism was originally written at zero
temperature. Then it was extended to finite temperature by
Amelino-Camelia and Pi for investigations of the effective
potential of the $\l\phi^4$ theory~\cite{ref16}.

In our following discussions of linear sigma model at finite
temperature, we use the imaginary time formalism, which is also
known as the Matsubara formalism~\cite{ref17}. This means
\bea\label{sum}
\int\0{d^4k}{(2\pi)^4}f(k)\to\01{\b}\sum_n\int\0{d^3\bf
k}{(2\pi)^3}f(i\o_n,{\bf k})  \equiv\int_\b f(i\o_n, {\bf k}),
\eea where $\b$ is the inverse temperature, $\b=1/T$; the
integration over the time component $k_0$ has been replaced by a
summation over discrete frequencies. For boson there are
$\o_n=2\pi nT$ and $n=0, \pm1, \pm2, \cdots$.  For the sake of
simplicity in what follows, a shorthand notation $\int_\b$ is used
to denote the integration and the summation.

\begin{figure}[tbh]
\begin{center}
\includegraphics[width=150pt,height=130pt]{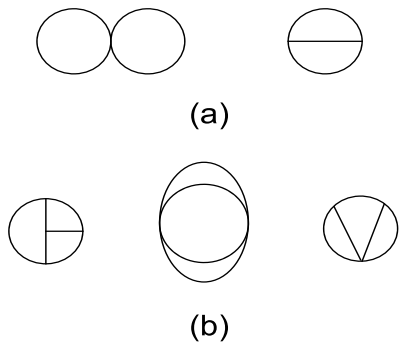}
\end{center}
\caption{Two-particle irreducible graphs which contribute to the
effective potential of a $\l\phi^4$ theory in the CJT formalism up
to (a) two loop level and (b) three loop level. The solid line
represents the full propagator.}\label{f1}
\end{figure}

For $\l\phi^4$ theory, according to the CJT
formalism~\cite{ref16}, the effective potential can be written as
\bea V(\phi,G)=U(\phi)+\012\int_\b\ln
G^{-1}(\phi;k)+\012\int_\b[D^{-1}(\phi;k)G(\phi;k)-1]+V_2(\phi,G),
\eea where $D$ and $G$ are the bare and full propagators of the
shifted $\l\phi^4$ theory respectively. The last term
$V_2(\phi,G)$ represents the sum of all two and higher-order loop
two-particle irreducible graphs of the theory with vertices given
by the interaction Lagrangian and propagators set equal to
$G(\phi;k)$. The diagram contribution to $V_2(\phi,G)$ for
$\l\phi^4$ theory are shown in figure \ref{f1}. $\phi$ and $G$
will be self-consistently determined by equation (\ref{mini}). In
the case of $\l\phi^4$ theory, the first diagram of figure
\ref{f1}(a) is the leading order in $V_2(\phi,G)$ in both the loop
expansion and the $1/N$ expansion. In our later discussion of
linear sigma model at Hartree and large $N$ approximation, one
needs to take into account the ``$\infty$" type diagram only.

\section{BEC at Hartree approximation}
From the Lagrangian (\ref{lag}), after shifting the sigma field we
can write down the tree level inverse propagators of $\s$, $\pi_0$
and $\pi_\pm$ respectively as
\bea D^{-1}_\s&=&\o^2_n+{\bf k}^2+m^2+\0\l2\phi^2, \\
D^{-1}_0&=&\o^2_n+{\bf k}^2+m^2+\0\l6\phi^2, \\
D^{-1}&=&(\o_n+i\mu)^2+{\bf k}^2+m^2+\0\l6\phi^2.  \label{pi} \eea
The equation (\ref{pi}) represents the inverse propagator of
$\pi_+$ and $\pi_-$. As the summation in equation (\ref{sum}) is
symmetric over $n$ from $-\infty$ to $+\infty$, $\o_n+i\mu$ and
$\o_n-i\mu$ are equivalent in describing the propagators of
$\pi_+$ and $\pi_-$. According to the CJT formalism, the effective
potential at finite temperature and finite isospin chemical
potential can be written as, \bea &&V(\phi,
G)=U(\phi)+\012\int_\b\ln G^{-1}_\s +\012\int_\b \left[D^{-1}_\s
G_\s-1\right]\nnu
\\ &+&\012\int_\b\ln G^{-1}_0
+\012\int_\b\left[D^{-1}_0G_0-1\right] +\int_\b\ln G^{-1}
+\int_\b\left[D^{-1}G-1\right]+V_2(\phi, G), \eea where $G_\s,
G_0$ and $G$ are the full propagators of $\s, \pi_0$ and $\pi_\pm$
respectively. They are determined by the stationary condition
(\ref{mini}). $V_2(\phi,G)$ represents the infinite sum of the two
particle irreducible vacuum graphs. However, at the Hartree
approximation we need only to calculate the ``$\infty$" (or
``double bubble") diagrams and treat each loop line as the full
propagator~\cite{ref10,ref15}. Therefore, $V_2$ can be written as
\bea V_2(\phi, G)&=&\0\l8\left[\int_\b
G_\s\right]^2+\0\l3\left[\int_\b
G\right]^2+\0\l8\left[\int_\b G_0\right]^2 \nnu \\
&+&\0\l6\int_\b G_\s\int_\b G+\0\l{12}\int_\b G_\s\int_\b G_0
+\0\l6\int_\b G\int_\b G_0. \eea

For the full propagators we could take the following ansatz,
\bea G^{-1}_\s&=&\o^2_n+{\bf k}^2+M^2_\s, \\
G^{-1}_0&=&\o^2_n+{\bf k}^2+M^2_0, \\ G^{-1}&=&(\o_n+i\mu)^2+{\bf
k}^2+M^2, \eea where the effective masses $M_\s, M_0$ and $M$ have
been introduced for $\s, \pi_0$ and $\pi_\pm$ respectively. As in
the Hartree approximation, only the tadpole diagrams contribute to
the self-energies. The effective masses are independent of
momentum. From the stationary condition (\ref{mini}), we obtain a
set of effective mass gap equations, \bea
M^2_\s&=&m^2+\0\l2\phi^2+\0\l2\int_\b
G_\s+\0\l3\int_\b G+\0\l6\int_\b G_0, \label{a1} \\
M^2_0&=&m^2+\0\l6\phi^2+\0\l2\int_\b G_0+\0\l6\int_\b
G_\s+\0\l3\int_\b G, \\ M^2&=&m^2+\0\l6\phi^2+\0{2\l}3\int_\b
G+\0\l6\int_\b G_\s+\0\l6\int_\b G_0. \label{a2} \eea Accordingly
the effective potential can be written as \bea V(\phi,
M)&=&U(\phi)+\012\int_\b\ln
G^{-1}_\s \nnu \\
&-&\012\int_\b(M^2_\s-m^2-\0\l2\phi^2)G_\s+\012\int_\b\ln G^{-1}_0
\nnu \\ &-&\012\int_\b(M^2_0-m^2-\0\l6\phi^2)G_0+\int_\b\ln G^{-1}
\nnu \\ &-&\int_\b(M^2-m^2-\0\l6\phi^2)G+V_2(\phi, M). \eea By
minimizing the potential with respect to the order parameter
$\phi$, we obtain one more equation, \bea
\left[m^2+\0\l6\phi^2+\0{\l}2\int_\b G_\s+\0{\l}6\int_\b
G_0+\0{\l}3\int_\b G\right]\phi-\e=0. \label{a3} \eea From
equations (\ref{a1})---(\ref{a2}) and (\ref{a3}), the effective
masses and order parameter can be solved self-consistently at
given temperature and chemical potential. There are four types of
integrals in above equations. After performing the Matsubara
frequency sums they are \bea Q(M)&=&\int_\b\ln[(\o_n+i\mu)^2+{\bf
k}^2+M^2] \nnu
\\ &=&Q_0(M)+Q_\b(M) \nnu \\
&=&\01{\b}\int\0{d^3{\bf
k}}{(2\pi)^3}\0{\o}2+\01{\b}\int\0{d^3{\bf
k}}{(2\pi)^3}\left[\ln(1-e^{-\b(\o+\mu)})+\ln(1-e^{-\b(\o-\mu)})\right],
\\ F(M)&=&\int_\b\01{(\o_n+i\mu)^2+{\bf k}^2+M^2} \nnu \\ &=&F_0(M)+F_\b(M) \nnu \\ &=&\int\0{d^3{\bf
k}}{(2\pi)^3}\01{2\o}+\int\0{d^3{\bf
k}}{(2\pi)^3}\01{2\o}\left[\01{e^{\b(\o+\mu)}-1}+\01{e^{\b(\o-\mu)}-1}\right], \\
Q(S)&=&\int_\b\ln(\o_n^2+{\bf k}^2+S^2) \nnu
\\ &=&Q_0(S)+Q_\b(S) \nnu \\
&=&\01{\b}\int\0{d^3{\bf k}}{(2\pi)^3}\0{\o_{\bf
k}}2+\02{\b}\int\0{d^3{\bf k}}{(2\pi)^3}\ln(1-e^{-\b\o_{\bf k}}),
\\ F(S)&=&\int_\b\01{\o_n^2+{\bf k}^2+S^2} \nnu \\ &=&F_0(S)+F_\b(S) \nnu \\ &=&\int\0{d^3{\bf
k}}{(2\pi)^3}\01{2\o_{\bf k}}+\int\0{d^3{\bf
k}}{(2\pi)^3}\01{\o_{\bf k}}\01{e^{\b\o_{\bf k}}-1}, \eea where
$\o=\sqrt{{\bf k}^2+M^2}$, $\o_{\bf k}=\sqrt{{\bf k}^2+S^2}$ and
$S=M_\s, M_0$. Each integral is divided into two parts: the
zero-temperature part which is divergent and the finite-
temperature part which is finite. The evaluation of the integral
requires renormalization. There are some discussions concerning
the renormalization on the CJT formalism in different
models~\cite{ref16,ref18,ref19}. The investigations about the
renormalization of the $O(N)$ linear sigma model can be referred
to~\cite{ref11,ref20}. In our discussion, we only keep the finite
temperature parts ($Q_\b$ and $F_\b$). The divergent parts of the
integrals are neglected as is done in~\cite{ref10,ref13,ref14}.

In the discussion of the thermodynamic system, the effective
potential $V$ is equivalent to the thermodynamic potential
$\Omega$. Thus we have \bea
\Omega&=&U(\phi)+\012Q_\b(M_\s)-\012(M^2_\s-m^2-\0\l2\phi^2)F_\b(M_\s)
\nnu \\ &+&\012Q_\b(M_0) -\012(M^2_0-m^2-\0\l6\phi^2)
F_\b(M_0)+Q_\b(M)-(M^2-m^2-\0\l6\phi^2)F_\b(M)+\Omega_2, \eea
where \bea
\Omega_2&=&\0\l8\left[F_\b(M_\s)\right]^2+\0\l3\left[F_\b(M)\right]^2+\0\l8\left[F_\b(M_0)\right]^2 \nnu \\
&+&\0\l6F_\b(M_\s)F_\b(M)+\0\l{12}F_\b(M_\s)F_\b(M_0)
+\0\l6F_\b(M)F_\b(M_0). \eea According to the relation \bea
\rho=-\0{\pl\Omega}{\pl\mu}, \eea and equation (\ref{a2}), we can
get the net charge density \bea \r=\int\0{d^3{\bf
k}}{(2\pi)^3}\left[\01{e^{\b(\o-\mu)}-1}-\01{e^{\b(\o+\mu)}-1}\right].
\label{a4} \eea This expression of density seems very similar to
that of ideal gas, but actually they are different. Because here
the effective mass $M$ in $\o$ is a function of temperature and
will be determined self-consistently by the gap equations.

Now we are in a position to discuss BEC. If we lower the
temperature of the system from a high temperature, it is known
that BEC will occur at the place where $\mu=M$ is reached. When
BEC occurs, equation (\ref{a4}) should be written as
\bea &&\r=\r_0+\r^*(\b,\mu=M), \nnu \\
&&\r^*(\b,\mu=M)=\int\0{d^3{\bf
k}}{(2\pi)^3}\left[\01{e^{\b(\o-M)}-1}-\01{e^{\b(\o+M)}-1}\right],
\label{a5} \eea where $\r_0$ represents the charge density of the
zero-momentum state~\cite{ref17}.
\subsection{Chiral Limit $\e=0$}
\begin{figure}[tbh]
\begin{center}
\includegraphics[width=220pt,height=130pt]{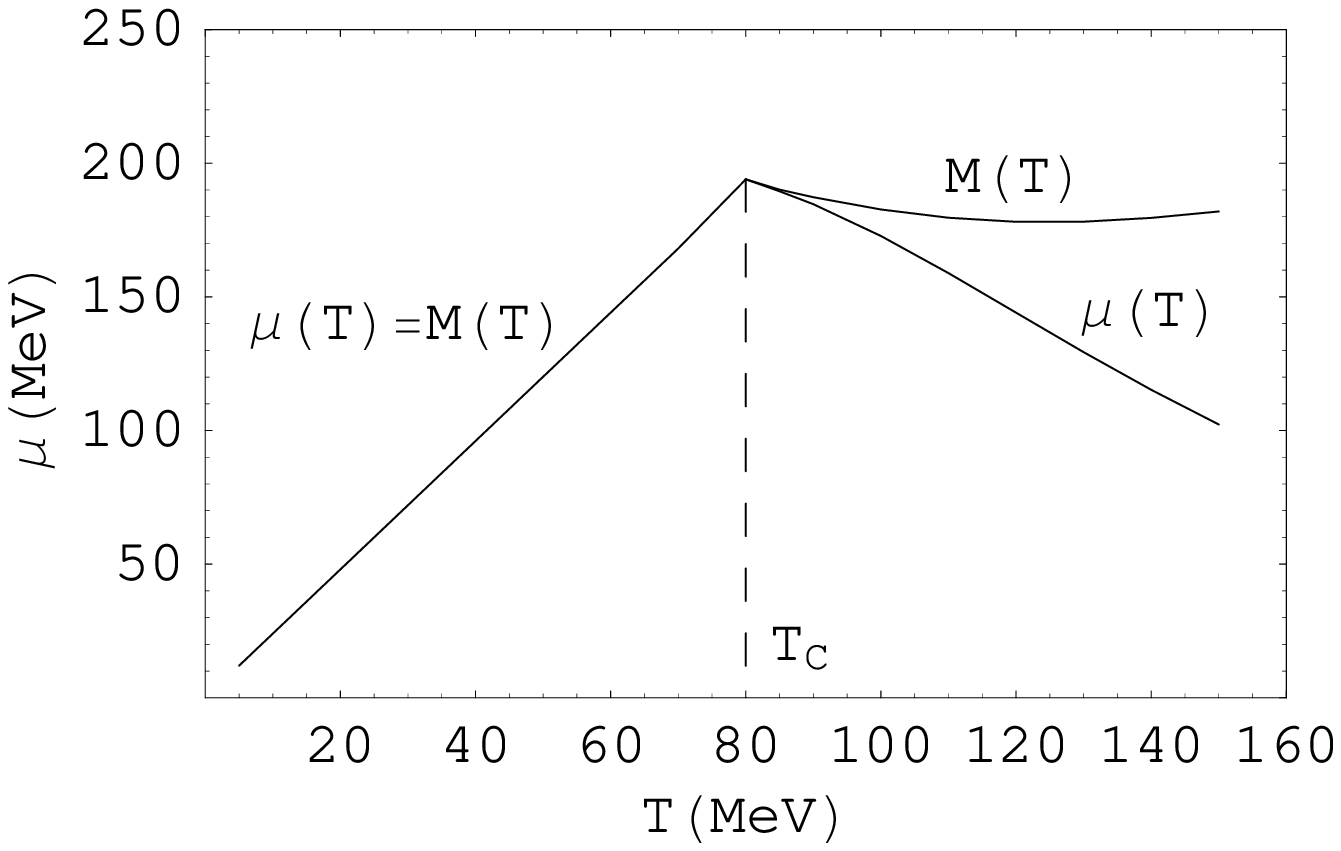}
\end{center}
\caption{$\mu$ and $M$ as functions of $T$ at the fixed total
charge density $\r=0.06fm^{-3}$. BEC happens at
$T_c=80MeV$.}\label{f2}
\end{figure}

In the chiral limit ($\e=0$), the corresponding coupling constant
and the negative mass parameter are given by $\l\approx 125$ and
$-m^2\approx 1.8\times 10^5MeV^2$. From equation (\ref{a4}), when
$\r$ is fixed, by solving the gap equations
(\ref{a1})---(\ref{a2}) and equation (\ref{a3}), we find both
$\mu$ and $M$ are functions of $T$. It can be plotted in figure
\ref{f2}. We can see that with $T$ decreasing, $M$ decreases first
and then increases, while $\mu$ keeps increasing quickly and
approaches $M$. At certain temperature, $\mu$ catches up with $M$
and BEC happens. From equation (31), we know the critical
temperature $T_c$ is determined implicitly by the equation \bea
\r=\r^*(\b_c,\mu=M). \eea When $T\leq T_c$, the system goes into
BEC phase. The equation (\ref{a5}) will be solved together with
the gap equations with $\r$ fixed. We find $\mu$ and $M$ are still
functions of $T$ and $\mu(T)=M(T)$. They both decrease with
temperature decreasing, which is indicated in figure \ref{f2}.

\begin{figure}[tbh]
\begin{center}
\includegraphics[width=220pt,height=130pt]{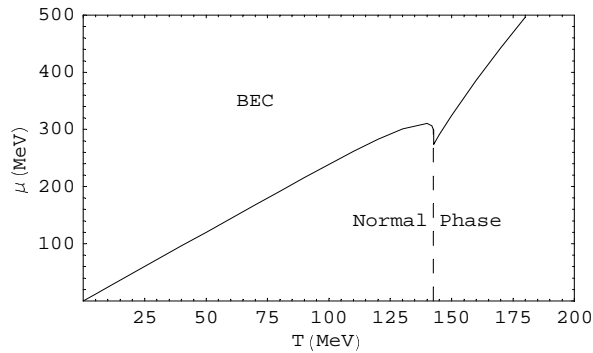}
\end{center}
\caption{The phase diagram of $\mu$ versus $T$ for BEC in the
chiral limit at Hartree approximation.}\label{f3}
\end{figure}

For different fixed total density $\r$, there will be different
values of critical temperature $T_c$ and chemical potential
$\mu(T_c)$, so one can study the $\mu-T$ phase diagram of BEC. It
is shown in figure \ref{f3}. In the chiral limit, at $T=0$, the
pion mass $M=0$, so the curve starts from the point $T=0$ and
$\mu=0$. There is a disjunction at $T\approx 143MeV$. This is due
to the first order chiral phase transition in the chiral limit at
Hartree approximation. At low temperature $\phi\neq 0$; at certain
high temperature $\phi=0$. From $\phi\neq 0$ to $\phi=0$ is
discontinuous, which is reflected in the BEC phase diagram as a
disjunction from low temperature to high temperature. Thorough
discussion about the relation of the BEC and the chiral phase
transition in the chiral limit can be referred to our previous
work~\cite{ref13}. In the chiral symmetry broken state ($\phi\neq
0$), the pion mass is non-zero ($\mu=M\neq 0$) which means the
Nambu-Goldstone(NG) theorem is not observed at Hartree
approximation as also indicated in the previous
literatures~\cite{ref10,ref21,ref22}. Recent discussions of this
direction can be referred to~\cite{ref23,ref24}. In this paper, we
mainly discuss BEC at the usually Hartree approximation and large
$N$ approximation. In our later discussion, we will see that at
large $N$ approximation the NG theorem will be preserved. In
figure \ref{f3}, we can see clear that the whole phase plane has
been divided into the BEC phase and the normal phase.

\subsection{Non-Chiral Limit $\e\neq 0$}
\begin{figure}[tbh]
\begin{center}
\includegraphics[width=220pt,height=130pt]{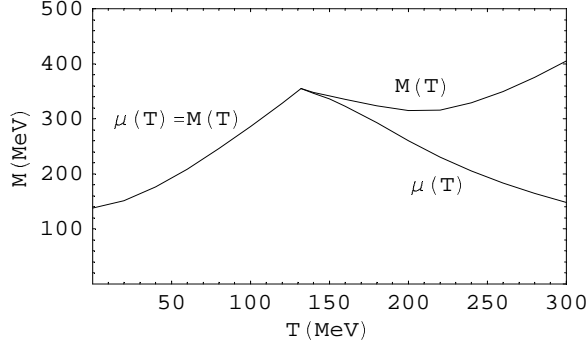}
\end{center}
\caption{$\mu$ and $M$ as functions of $T$ at the fixed total
charge density $\r=0.3fm^{-3}$. BEC happens at
$T_c=132MeV$.}\label{f4}
\end{figure}

\begin{figure}[tbh]
\begin{center}
\includegraphics[width=220pt,height=130pt]{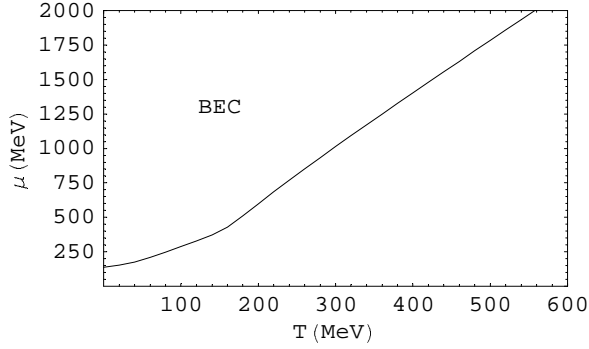}
\end{center}
\caption{The phase diagram of $\mu$ versus $T$ for BEC in the
non-chiral limit at Hartree approximation.}\label{f5}
\end{figure}

In the non-chiral limit ($\e\neq 0$), the corresponding coupling
constant, the negative mass parameter and the explicit symmetry
breaking term are given by $\l\approx 1.18\times 10^2$,
$-m^2\approx 1.51\times 10^5MeV^2$ and $\e\approx 1.77\times
10^6MeV^3$. When equation (\ref{a4}) is solved together with the
gap equations (\ref{a1})---(\ref{a2}) and equation (\ref{a3}) at
fixed $\r$, the critical temperature $T_c$ of BEC will be
determined at $\mu=M$ which is shown in figure \ref{f4}.
 The procedure is similar to that of
the chiral limit. For different fixed $\r$, $\mu$ varies with
$T_c$, so the $\mu-T$ phase diagram of BEC in the non-chiral limit
can be plotted as shown in figure \ref{f5}. At zero temperature,
the vacuum mass of pion is $138MeV$, so the curve starts from the
value of $\mu_c=m_\pi=138MeV$. The chiral phase transition in the
non-chiral limit is a smooth crossover. There is no discontinuity
in the BEC phase diagram.
\section{BEC at large $N$ approximation}
The generalized version of the meson sector of the linear sigma
model is called $O(N)$ model and is based on a set of $N$ real
scalar fields. The $O(N)$ model Lagrangian can be written as \bea
{\cal L}=\012(\pl\bm{\Phi})^2-\012m^2\bm{\Phi}^2-
\01{6N}\l\bm{\Phi}^2+\e\s , \eea where $\bm{\Phi}$ can be
identified as $\bm{\Phi}=(\s, \pi_1, \pi_2, \cdots, \pi_{N-1})$.
The last term is the symmetry breaking term in order to generate
the observed masses of the pions. If $N=4$, it becomes the $O(4)$
linear sigma which has been discussed above.

For the $O(N)$ linear sigma model, after introducing the chemical
potential and shifting the sigma field the Lagrangian can be
written as \bea {\cal
L}&=&(D_\mu\pi)^+(D^\mu\pi)^--(m^2+\0{2\l}{3N}\phi^2)|\pi|^2+\012\sum_{i}(\pl\pi_i)^2+\012(\pl\s)^2
 \nnu \\
&-&\012(m^2+\0{2\l}{3N}\phi^2)\sum_{i}\pi^2_i-\012(m^2+\0{2\l}N\phi^2)\s^2-\012m^2\phi^2-\0\l{6N}\phi^4+\e\phi
\nnu \\
&-&\0\l{6N}\sum_{i,j}\left[\s^4+4|\pi|^4+4\s^2|\pi|^2+2\s^2\pi_i^2+4|\pi|^2\pi_i^2+\pi_i^4+2\pi_i^2\pi_j^2\right]
\eea where $|\pi|^2=\pi_+\pi_-$, the sum over $i$ or $j$ is from
$3$ to $N-1$ and $i\neq j$. The chemical potential here is
associated with the conserved charge of an $O(2)$
symmetry~\cite{ref25}. Then we can write down the tree level
inverse propagators of $\s$, $\pi_i$ and $\pi_\pm$
respectively as \bea D^{-1}_\s&=&\o^2_n+{\bf k}^2+m^2+\0{2\l}N\phi^2,  \\
D^{-1}_i&=&\o^2_n+{\bf k}^2+m^2+\0{2\l}{3N}\phi^2, \\
D^{-1}&=&(\o_n+i\mu)^2+{\bf k}^2+m^2+\0{2\l}{3N}\phi^2. \eea From
the CJT formalism, the thermodynamic potential of the $O(N)$
linear sigma model can be written as \bea &&\Omega(\phi,
G)=\012m^2\phi^2+\0\l{6N}\phi^4-\e\phi+\012\int_\b\ln G^{-1}_\s
+\012\int_\b \left[D^{-1}_\s G_\s-1\right]\nnu
\\ &+&\0{N-3}2\int_\b\ln G^{-1}_i
+\0{N-3}2\int_\b\left[D^{-1}_iG_i-1\right] +\int_\b\ln G^{-1}
+\int_\b\left[D^{-1}G-1\right]+\Omega_2(\phi, G), \eea where \bea
\Omega_2(\phi, G)&=&\0\l{2N}\left[\int_\b
G_\s\right]^2+\0{4\l}{3N}\left[\int_\b
G\right]^2+\0{(N-1)(N-3)\l}{6N}\left[\int_\b G_i\right]^2 \nnu \\
&+&\0{2\l}{3N}\int_\b G_\s\int_\b G+\0{(N-3)\l}{3N}\int_\b
G_\s\int_\b G_i +\0{2(N-3)\l}{3N}\int_\b G\int_\b G_i. \eea

By minimizing the thermodynamic potential with the full
propagators, we obtain the following set of effective mass gap
equations \bea\label{eq1}
M^2_\s&=&m^2+\0{2\l}N\phi^2+\0{2\l}N\int_\b
G_\s+\0{4\l}{3N}\int_\b G+\0{2(N-3)\l}{3N}\int_\b G_i, \\
M^2_i&=&m^2+\0{2\l}{3N}\phi^2+\0{2(N-1)\l}{3N}\int_\b
G_i+\0{2\l}{3N}\int_\b G_\s+\0{4\l}{3N}\int_\b G, \\
M^2&=&m^2+\0{2\l}{3N}\phi^2+\0{8\l}{3N}\int_\b
G+\0{2\l}{3N}\int_\b G_\s+\0{2(N-3)\l}{3N}\int_\b G_i, \eea where
$M_i$ stands for the effective mass of $\pi_i$ ($i=3,\cdots,N-1$).
By minimizing the thermodynamic potential with the order parameter
$\phi$, we have one more equation \bea\label{eq2}
\left[m^2+\0{2\l}{3N}\phi^2+\0{2\l}N\int_\b
G_\s+\0{2(N-3)\l}{3N}\int_\b G_i+\0{4\l}{3N}\int_\b
G\right]\phi-\e=0. \eea

In the large $N$ approximation, which means that we ignore the
terms of $O(1/N)$, the equations $(\ref{eq1})-(\ref{eq2})$ reduce
to \bea M^2_\s=m^2+\0{2\l}N\phi^2+\0{2\l}{3}F_\b (M_i), \\
M^2_i=m^2+\0{2\l}{3N}\phi^2+\0{2\l}{3}F_\b (M_i), \label{eq3} \\
M^2=m^2+\0{2\l}{3N}\phi^2+\0{2\l}{3}F_\b (M_i). \label{eq4} \\
\left[m^2+\0{2\l}{3N}\phi^2+\0{2\l}{3}F_\b (M_i)\right]\phi-\e=0.
\label{eq5} \eea The terms quadratic in $\phi$ are not of $O(1/N)$
but of $O(1)$ since $\phi$ depends on $N$ as $\phi^2=3Nm^2/2\l$.
For the integration only the finite temperature part is preserved
as done in Hartree approximation. The effective mass equations
(\ref{eq3}) and (\ref{eq4}) are identical which shows the
effective mass of $\pi_\pm$ is equal to that of $\pi_i$.

From the thermodynamic potential we can also derive the net charge
density as \bea \r=\int\0{d^3{\bf
k}}{(2\pi)^3}\left[\01{e^{\b(\o-\mu)}-1}-\01{e^{\b(\o+\mu)}-1}\right],
\label{eq6} \eea where $\o=\sqrt{{\bf k}^2+M^2}$. In the following
discussion we will set $N=4$ for our numerical evaluation.
\subsection{Chiral Limit $\e=0$}
In the chiral limit, by the equation (\ref{eq4}), the equation
(\ref{eq5}) could be written as \bea \phi M^2=0. \eea When
$\phi\ne 0$, the effective mass of pion will be zero.
\begin{figure}[tbh]
\begin{center}
\includegraphics[width=220pt,height=130pt]{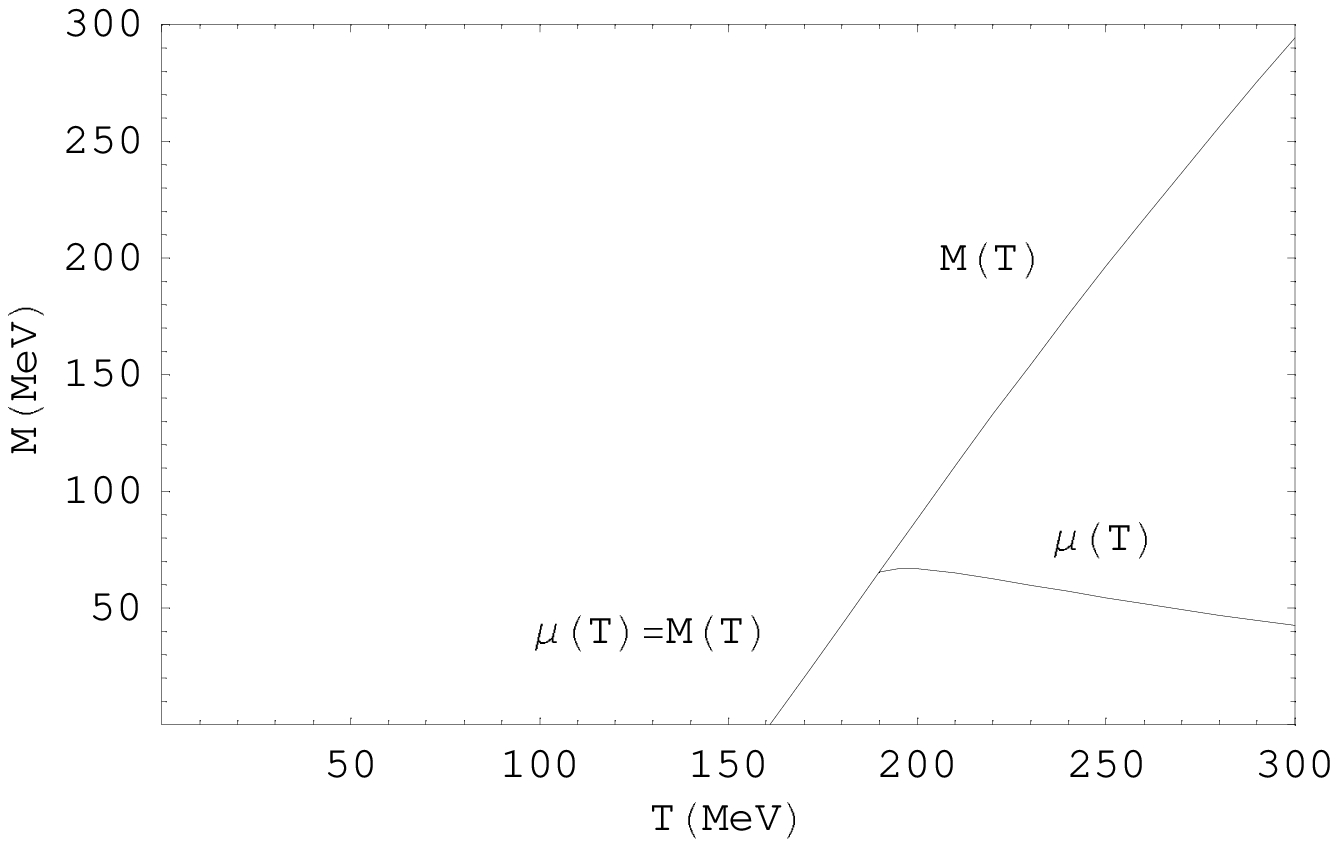}
\end{center}
\caption{$\mu$ and $M$ as functions of $T$ at the fixed total
charge density $\r=0.1fm^{-3}$. BEC happens at
$T_c=190MeV$.}\label{f6}
\end{figure} Therefore the pions are massless
Goldstone bosons in the chiral symmetry broken phase which means
the NG theorem is observed in the large $N$ approximation. In
chiral symmetry broken phase BEC will happen at $\mu=M=0$. At
certain high temperature that the chiral symmetry restored and
$\phi=0$, pion becomes massive because of the thermal contribution
to the effective mass. In this case the equation (\ref{eq4}) and
(\ref{eq6}) will be solved together at fixed density to determine
the critical temperature of BEC as shown in figure \ref{f6}.
\begin{figure}[tbh]
\begin{center}
\includegraphics[width=220pt,height=130pt]{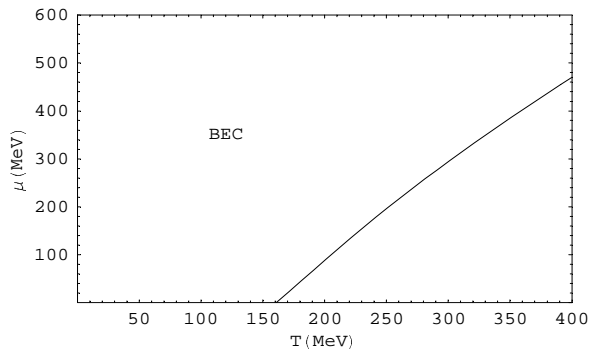}
\end{center}
\caption{The phase diagram of $\mu$ versus $T$ for BEC in the
chiral limit at large $N$ approximation.}\label{f7}
\end{figure} For different density by
determining the critical temperature and chemical potential, we
can plot The $\mu-T$ phase diagram of BEC as show in figure
\ref{f7}. At low temperature when chiral symmetry is not restored,
the BEC will happen at $\mu(T)=M(T)=0$ which reflects the
requirements of the NG theorem; At high temperature when chiral
symmetry is restored, the BEC happens when $\mu$ equal to the
nonzero effective mass of pion.

\subsection{Non-Chiral Limit $\e\neq 0$}
\begin{figure}[tbh]\begin{center}
\includegraphics[width=220pt,height=130pt]{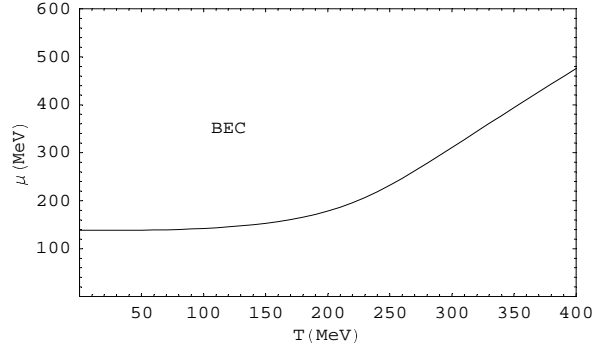}
\end{center}
\caption{The phase diagram of $\mu$ versus $T$ for BEC in the
non-chiral limit at large $N$ approximation.}\label{f9}
\end{figure}

In the non-chiral limit, the equation (\ref{eq4}), (\ref{eq5}) and
(\ref{eq6}) will be solved together at fixed density to find the
critical temperature of BEC.
\begin{figure}[tbh]
\begin{center}
\includegraphics[width=220pt,height=130pt]{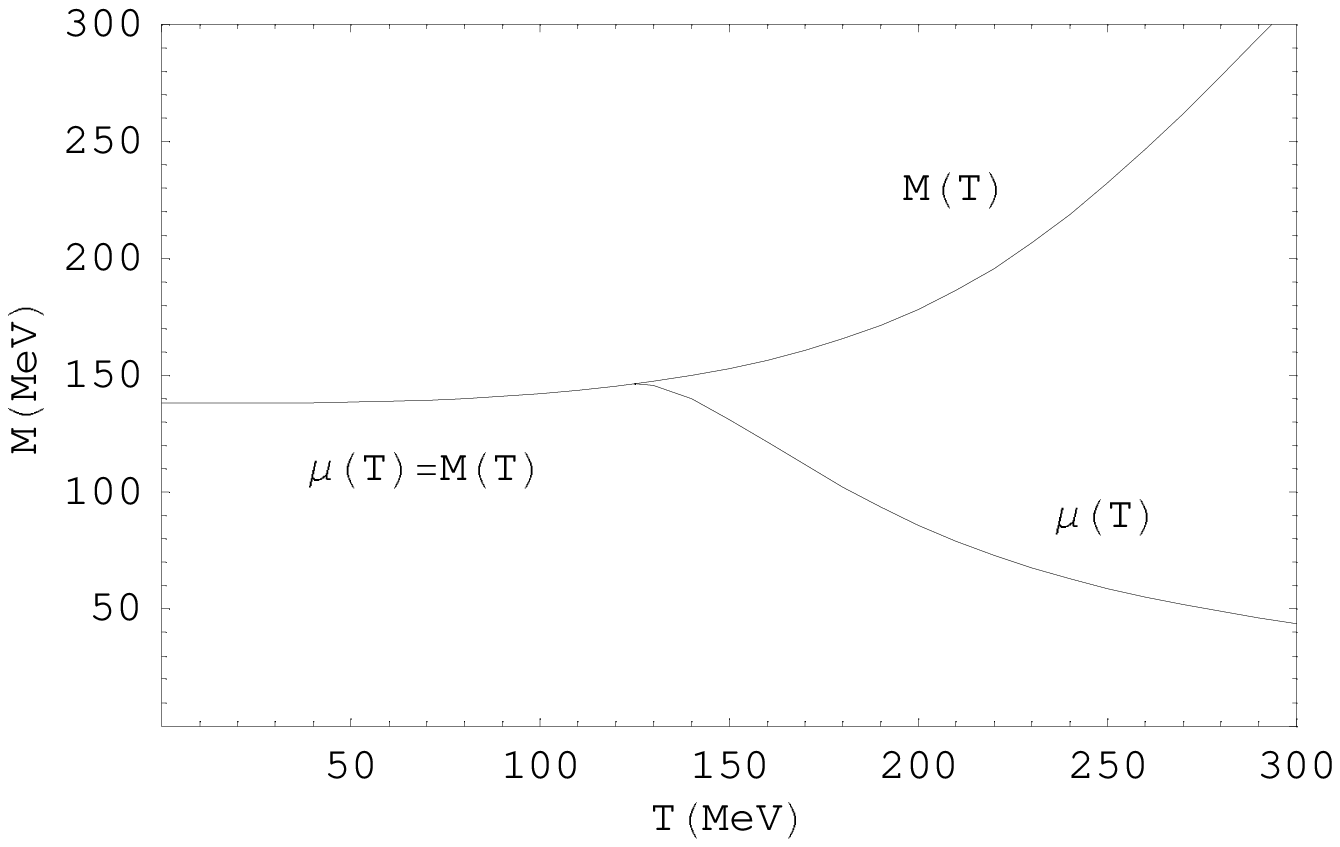}
\end{center}
\caption{$\mu$ and $M$ as functions of $T$ at the fixed total
charge density $\r=0.1fm^{-3}$. BEC happens at
$T_c=125MeV$.}\label{f8}
\end{figure}
The figure \ref{f8} shows the critical temperature at the fixed
density is determined just at the time $\mu(T)=M(T)$. By the same
procedure as that in the chiral limit, the $\mu-T$ phase diagram
could be plotted as in figure \ref{f9}.
 At zero temperature the vacuum mass of pion
is $138MeV$, and the BEC happens at critical chemical potential
$\mu_c=138MeV$. When temperature increases the critical chemical
potential also increases. The BEC phase is in the upper plane of
the phase diagram.

\section{summary}
By the CJT formalism we have derived the temperature and density
dependent effective potential based on the linear sigma model. The
BEC is investigated at the Hartree approximation and the large $N$
approximation. The critical temperature of BEC is determined by
lowering the temperature of the system at the fixed density to
find the critical point at which $\mu(T)=M(T)$. The $\mu-T$ phase
diagram of BEC has been plotted in different situations. The main
features of BEC at different approximations can be summarized as
below:

(1) At Hartree approximation and in chiral limit, for the $\mu-T$
phase diagram of BEC, the phase separate line is discontinuous
from low temperature to high temperature. At low temperature phase
($\phi\ne0$), BEC happens at $\mu=M\geq 0$, which shows the NG
theorem is not observed.

(2)At Hartree approximation and in non- chiral limit, the critical
chemical potential of BEC at zero temperature is exactly the
vacuum mass of pion ($138MeV$). With temperature increasing the
critical chemical potential increases continuously.

(3) At large $N$ approximation and in chiral limit, at low
temperature phase before chiral symmetry restored ($\phi\ne 0$),
BEC happens exactly at $\mu(T)=M(T)=0$, which shows the NG theorem
is observed. Above certain temperature when chiral symmetry is
restored, the critical chemical potential of BEC becomes non-zero
and increases with temperature increasing.

(4) At large $N$ approximation and in non-chiral limit, the
critical chemical potential of BEC is $138MeV$ at zero temperature
then increases with temperature increasing.

\begin{acknowledgments}
This work was supported in part by the National Natural Science
Foundation of China with No. 10547112 and No. 10675052.
\end{acknowledgments}

\end{document}